\documentclass[fleqn,10pt]{wlscirep}
\usepackage{amsmath}
\usepackage{algorithm}
\usepackage[noend]{algpseudocode}
\usepackage[colorinlistoftodos]{todonotes}
\usepackage{graphicx,graphics,subfigure,multicol} 
\usepackage{url}
\usepackage{booktabs}
\usepackage{multirow}
\usepackage{threeparttable}
\usepackage{tabularx} 
\usepackage{todonotes}
\usepackage{ulem}
\usepackage{lscape}
\usepackage{wasysym}

\makeatletter
\g@addto@macro\appendix{\setcounter{figure}{0}}
\g@addto@macro\appendix{\setcounter{table}{0}}
\makeatother
\title{Periodicity in Movement Patterns Shapes Epidemic Risk in Urban Environments}

\author[1,*]{Zhanwei Du}
\author[1]{Spencer J Fox}
\author[2]{Petter Holme}
\author[3]{Jiming Liu}
\author[4,5]{Alison P. Galvani}
\author[1,6]{Lauren Ancel Meyers}

\affil[1]{Department of Integrative Biology, University of Texas at Austin, Austin, TX, USA}
\affil[2]{Institute of Innovative Research, Tokyo Institute of Technology, Yokohama, Japan}
\affil[3]{Department of Computer Science, Hong Kong Baptist University, Kowloon Tong, Hong Kong}
\affil[4]{Center for Infectious Disease Modeling and Analysis, Yale School of Public Health, New Haven, CT, USA}
\affil[5]{Department of Ecology and Evolution, Yale University, New Haven, CT, USA}
\affil[6]{Santa Fe Institute, Santa Fe, NM, USA}

\affil[*]{duzhanwei0@gmail.com}

\begin{abstract}
Daily variation in human mobility modulates the speed and severity of emerging outbreaks , yet most epidemiological studies assume static contact patterns.
With a highly mobile population exceeding 24 million people, Shanghai, China is a transportation hub at high risk for the importation and subsequent global propagation of infectious diseases. 
Here, we use a dynamic metapopulation model informed by hourly transit data for Shanghai to estimate epidemic risks across thousands of outbreak scenarios. We find that the rate of initial epidemic growth varies by more than twenty-fold, depending on the hour and neighborhood of disease introduction. 
The riskiest introductions are those occurring close to the city center and on Fridays--which bridge weekday and weekend transit patterns and thereby connect otherwise disconnected portions of the population.
The identification of these spatio-temporal hotspots can inform more efficient targets for sentinel surveillance and strategies for mitigating transmission.
\end{abstract}

\begin{document}
\flushbottom
\maketitle
\thispagestyle{empty}

\section*{Introduction}
Despite advances in medical technology, emerging infectious diseases threaten global health and can exact devastating socioeconomic repercussions. Urban settings are prone to outbreaks, as they often serve as transportation hubs and have high population densities. The rate of dissemination and the effectiveness of control efforts are affected by the contact patterns underlying transmission, as demonstrated by the heterogeneous global spread of SARS in 2003-2004 \cite{dye2003modeling, riley2003transmission, meyers2005network} and the variable urban expansion of Ebola in West Africa in 2014-2015 \cite{team2014ebola,pandey2014strategies}. 
The  early emergence of the 2009 H1N1 pandemic also reflected the epidemiological importance of human movement, with the spring break holiday period both suppressing transmission within Mexico City \cite{dimitrov2009optimizing,pourbohloul2009initial} and exacerbating global movement via vacationers returning from Mexico \cite{fraser2009pandemic}.

As outbreaks emerge within cities, the spatiotemporal dynamics of transmission depend on physical encounters between people during their daily activities  \cite{merler2011determinants,xia2013identifying} 
Human mobility studies provide data on the spatiotemporal co-location of individuals in large populations \cite{dalziel2013human,eubank2004modelling}, which have been used to model epidemics, including the transmission of  2009 H1N1 \cite{balcan2009seasonal,dalziel2013human} and 2014 Ebola  \cite{merler2015spatiotemporal}  in urban settings \cite{cooley2011role,frias2011agent,dalziel2013human,hoen2015epidemic,herrera2016disease}.
These studies show that epidemic dynamics can vary substantially across locations and depend on interactions between intrinsic epidemiological characteristics of the spreading pathogens and human contact networks.
In particular, urban contact networks often consist of coherent communities--groups of people who are highly intra-connected, perhaps via physical neighborhood or common school, workplace, or religious organization, but only loosely interconnected \cite{girvan2002community,PhysRevE.70.066111}.
While local affiliations can amplify disease transmission, separation between communities can impede longer range spread \cite{salathe2010dynamics,hoen2015epidemic}.

Many mobility-based simulations of urban disease transmission capture geo-spatial heterogeneity in contact patterns, but assume that these social networks are static over the course of days and weeks \cite{hoen2015epidemic,zhu2016inferring}. However, urban contact patterns shift on time scales relevant to the emergence and control of outbreaks. They exhibit hourly and daily fluctuations that impact not only individual risk and pairwise associations, but larger scale community structure, with groups transiently assembling and dissolving as people go about their daily lives \cite{du2017understanding,zhong2015measuring,long2015finding,gong2015inferring}. 
The timing of a disease introduction can strongly influence the course of emergence, as seen in the seasonal or pandemic emergence of influenza \cite{cooley2016weekends,Fox2017td} and arboviruses like dengue \cite{robert2016modeling}. 
On a finer timescale, several recent studies found that transmission rates may decrease considerably during weekends and other school breaks \cite{towers2012impact, cooley2016weekends, eggo2016respiratory,cauchemez2008estimating,hens2009estimating}.

Here, we study the spread of a directly-transmitted disease introduced into a realistically dynamic urban network, at a fine spatiotemporal resolution. We hypothesize that the risk and magnitude of epidemic emergence will depend critically on the day of the week, daily period, and neighborhood of the introduction. 
We focus on Shanghai, China--a global financial center and transport hub at particularly at high risk for emerging or reemerging epidemics. For example, Shanghai reported the first confirmed case of H7N9 influenza during the 2013 outbreak~\cite{world2013overview}.
Using a metapopulation susceptible-infected-recovered model of disease transmission  based on subway transit patterns of over 11 million of the 24 million inhabitants of Shanghai, China in 2014 \cite{Bulletin2014}, we closely track the spread of a 2009 H1N1 pandemic-like influenza virus \cite{fraser2009pandemic}. We find that the highest risk scenarios include introductions occurring near the city center and on Fridays, with large numbers of initial cases and high reproduction numbers.

\section*{Methods}

\subsection*{Data} 
The Shanghai metropolitan subway system is one of the longest rapid transit system in the world \cite{unionMetro2015}.
We obtained one month of subway usage data (April 2015) from the Shanghai Public Transportation Card Co.\ Ltd, released by Shanghai Open Data Apps \cite{coltd2015}. It includes comprehensive records from the travel smart cards of approximately 11 million individuals, totaling over 120 million trips among Shanghai's 313 subway stations. Each trip record contains the check-in and check-out times and stations.

\subsection*{Construction of metapopulations}
We divide Shanghai into 313 neighborhoods (henceforth \textit{locations}), corresponding to the 313 subway stations. Shanghai has 15 administrative districts, each of which contains one or more subway stations. We assume that the population in any given district is equally divided across the locations within the district. That is, we set the population at the $i$-th location (in district $d$) equal to $N_i=\frac{N_d}{c_d}$ where $N_d$ denotes the total population of the $d$-th district according to the 2015 Shanghai statistical yearbook \cite{SSY2015} and $c_d$ is the number of locations within the $d$-th district. For example, in a district with a population of 1,000,000 and 10 subway stations, each station would be assumed to contain 100,000 people. 

The Shanghai subway system runs daily between 5:30 and 24:00. Following a prior study \cite{du2017understanding}, we divide each day into four periods of approximately five hours each: morning (5:30 to 09:59), noon (10:00 to 15:59), afternoon (16:00 to 20:59), and evening (21:00 to 23:59). The morning and afternoon periods contain the two major daily commuting peaks. To facilitate analysis, we consider only trips that begin and end in the same period, and analyze our simulation dynamics at the temporal scale of the period \cite{zhong2016variability}. 

To simulate human mobility within Shanghai, we construct a series of 168 flow matrices $G^t$ connecting the 313 locations, one for each hour of the week $t$, from which we derive 28 aggregate flow matrices $F^p$, one for each daily period $p$. These are estimated from subway movement records from week 16, which represented a typical week without holiday breaks. 
The $ij$-th entry in an hourly mobility matrix is given by $H_{ij}^t=M_{ij}^t/N_{i}$ where $M_{ij}^t$ is the number of travelers from location $i$ to location $j$ during hour $t$ and $N_{i}$ is the population size at location $i$. The corresponding entries in the aggregate mobility matrices are then given by $F_{ij}^p = \sum_{t\in p} H_{ij}^t$ for each time period $p$. 

\subsection*{Epidemic dynamics}
We model the spread of a directly transmitted disease within and between the 313 Shanghai locations. For each location $i$, the population is divided into three disease compartments---susceptible (individuals who can get the disease), infective (individuals who has the disease and can spread it further) and recovered (individuals who are immune and cannot spread the disease)---with the number of individuals in compartment $i$ at time $t$ denoted by $S_i(t)$, $I_i(t)$ and $R_i(t)$, respectively. 

We let $\beta$ and $\gamma$ denote the transmission and recovery rates of the disease, respectively, across all locations in Shanghai. The average infectious period is thus $1/\gamma$ and the local basic reproduction number (i.e., the expected number of secondary cases produced by a single infection in a fully susceptible and well-mixed population) is $R_0 = \beta/\gamma$. We use the following deterministic equations to simulate metapopulation susceptible–infectious–recovered (SIR) dynamics throughout the $p=313$ Shanghai neighborhoods: 
\begin{eqnarray}
{S_i}(t+1)-S_i(t) &=& -\beta S_i(t) I_i(t) /N_i + \sum_{j=1}^{p} F_{ji}^tS_j(t) - \sum_{j=1}^{p} F_{ij}^tS_i(t) \\
{I_i}(t+1)-{I_i}(t) &=& \beta S_i(t) I_i(t) /N_i + \sum_{j=1}^{p} F_{ji}^tI_j(t) - \sum_{j=1}^{p} F_{ij}^tI_i(t) - \gamma I_i(t) \\
{R_i}(t+1)-R_i(t) &=& \gamma I_i(t) + \sum_{j=1}^{p} F_{ji}^tR_j(t) - \sum_{j=1}^{p} F_{ij}^tR_i(t)
\end{eqnarray}
The first term in each equation governs movement through the susceptible and infective compartments and represents local disease transmission. The second and third terms represent the arrival and departure of individuals via public transit, respectively. We consider two sets of disease scenarios: a low influenza-like transmission rate ($R_0 = 1.5$), similar to that estimated for the 2009 H1N1 pandemic\cite{fraser2009pandemic}, and a more highly infectious disease with $R_0=7.5$. We assume an hourly recovery rate of $\gamma = 0.33/24$, and hourly transmission rates of $\beta=0.5/24$ for low $R_0$ and $\beta=2.33/24$ for high $R_0$. We analyze $313\cdot 3 \cdot 28 \cdot 2 =52,584$ 
different introduction scenarios: all combinations of (a) $313$ locations, (b) initial outbreak sizes, $I_{0}$ of 1, 100 or 10,000 (corresponding to a range of both natural and bioterrorism introduction scenarios \cite{masuda2017introduction}), (c) 28 different periods of the week (seven days, four periods per day), and (d) low or high $R_0$. 

We quantify the (metapopulation) epidemic growth rate by tracking the emergence of disease at new locations. Let $A(d,p)$ denote the number of locations (out of 313) that receive their first infection during period $p$ on day $d$. Then $A(d,p)/313$ gives the fraction of newly infected locations ($\phi$) during that period. 
Then the cumulative fraction of infected locations by day $d\ge 1$ is given by $\Phi(d) = \sum_{j=0}^{d}\sum_{p=1}^{4} A(j,p)/313$, since there are four periods per day. 

To compare the epidemic growth across different initial outbreak scenarios, we track time until $\Phi(d)$ reaches specified thresholds, such as $T=10\%$ and call this quantity $\Gamma_T$. For example, $\Gamma_{10\%}$ is the number of days following the initial introduction at which 10\% of the 313 locations have experienced at least one infection. To assess the epidemiological vulnerability of a specific location $l$, we track the days until $l$ becomes infected under various scenarios ($\chi$).

\subsection*{Shanghai contact network analysis}

We reinterpret the flow matrices ($F^p$) as adjacency matrices that describe the structure Shanghai's dynamic metapopulation contact network. For each period $p$, the vertices are the 313 locations and the edge between a given pair vertices, $i$ and $j$, is weighted by the flow during $p$ ($F_{ij}^p$). To  estimate correlations in mobility patterns across time periods, we calculate and test the statistical significance of Pearson product-moment correlation coefficients between flow matrices.

For each day of the week $d$, we estimate a quantity that we call \textit{network coherence} ($C_d$), which is related to the conventional concept of a network coherence (longest among the shortest paths connecting pairs of nodes) and summarizes the epidemiological extent of the population. Specifically, we use the weighted flow matrices ($F^{\text{morn}}$, $F^{\text{noon}}$, $F^{\text{aftn}}$, and $F^{\text{eve}}$) to solve for the highest probability path between every pair of locations in a given day, and then find the least likely of those high probability paths. We assume that individuals can move along at most one edge during each of the four time periods. Thus, a path from location $i$ to location $j$ during day $d$ is a chronological sequence of at most four moves, originating in $i$ and ending in $j$~\cite{holme2015modern}. 
Specifically, we find the shortest path from $i$ to $j$ during day $d$ ($P_{ij}^d$) by maximizing the product $F_{n_1,n_2}^{\text{morn}}\cdot F_{n_2,n_3}^{\text{noon}}\cdot F_{n_3,n_4}^{\text{aftn}}\cdot F_{n_4,n_5}^{\text{eve}}$ over all possible $1 \le {n_1,n_2,n_3,n_4,n_5}\le 313$ such that there exists an $a$ and $b$ where $1 \le a<b\le 5$, $n_a=i$, and $n_b=j$. 
The daily coherence of the Shanghai network ($C_d$) is then given by the minimum $P_{ij}^d$, over all possible pairs of locations, $i$ and $j$. It approximates the probability that somebody will travel between the most disconnected locations on day $d$.
The larger the coherence, the more readily travelers can transmit disease throughout the entire urban region. 
As an alternative statistic of transmission efficiency, we also take the average probability that an individual will travel between any two locations in Shanghai on each day $d$ (denoted $\delta_d$), over all possible pairs of locations.

We roughly distinguish travel for work versus non-work, assuming that the typical worker travels to work by 10:00 and leaves work after 17:00. This workday corresponds to traffic peaks in the Shanghai subway system~\cite{unionMetro2015}.
If the first origin station matches the last destination station of the day, we label it 'home' and  label the station visited for the longest period in between as 'work' and all other stations visited after 16:00 as 'non-work', which may include errands, health or beauty appointments, entertainment, etc. On a given day, we assume that a working individual will fall into one of six classes depending on the series of work and non-work locations visited (Fig.~\ref{figActivities}). We let $N^d_\text{W}$ be the number of workers on day $d$ and $N^d_\text{E}$ be the number of those workers that make extra non-work stops on day $d$. 
The {\it recreational fraction}, $\rho_d=N^d_\text{E}/N^d_\text{W}$, approximates the daily willingness to travel for non-work related activities.

To calculate the network centrality of individual locations, we consider the network defined by the  daily mobility matrix $\hat{F}$ averaged across the entire week, where $\hat{F}_{ij} = \frac{1}{7}\sum_{p} F_{ij}^p$. The in-degree and out-degree of location $i$ are then given by $k_\text{in}(i)=\sum_j \hat{F}_{ji}$ and $k_\text{out}(i)=\sum_j \hat{F}_{ij}$, respectively. 
The shortest path in  $\hat{F}$ from $i$ to $j$  ($\hat{P}_{ij}$) is defined as the path ($v_1, v_2, ..., v_n$), where $v_1=i$ and $v_n=j$, over all possible paths maximizes the product $\prod_{k=1}^{n-1} \hat{F}_{k,k+1}$.
The in-closeness centrality of location $i$ is the inverse sum of shortest path distances to the location from all other reachable locations in the network, and is given by $c_\text{in}(i)=-1/(\sum_j log(\hat{P}_{ij}))$. Likewise, the out-closeness centrality of a location $i$ is given by $c_\text{out}(i)=-1/(\sum_j log(\hat{P}_{ij}))$.

We analyze {\it community structure} dynamics across the 28 periods by partitioning each of the $p$ flow networks into modules using the Louvain community detection algorithm~\cite{blondel2008fast}. It identifies disjoint subsets of locations such that their intra-connectivity far exceeds their inter-connectivity. We then visualize the community dynamics throughout the week via MapEquation alluvial diagrams~\cite{edler2013mapequation}.

\begin{figure}[!h]
\centering \includegraphics[scale=0.3]{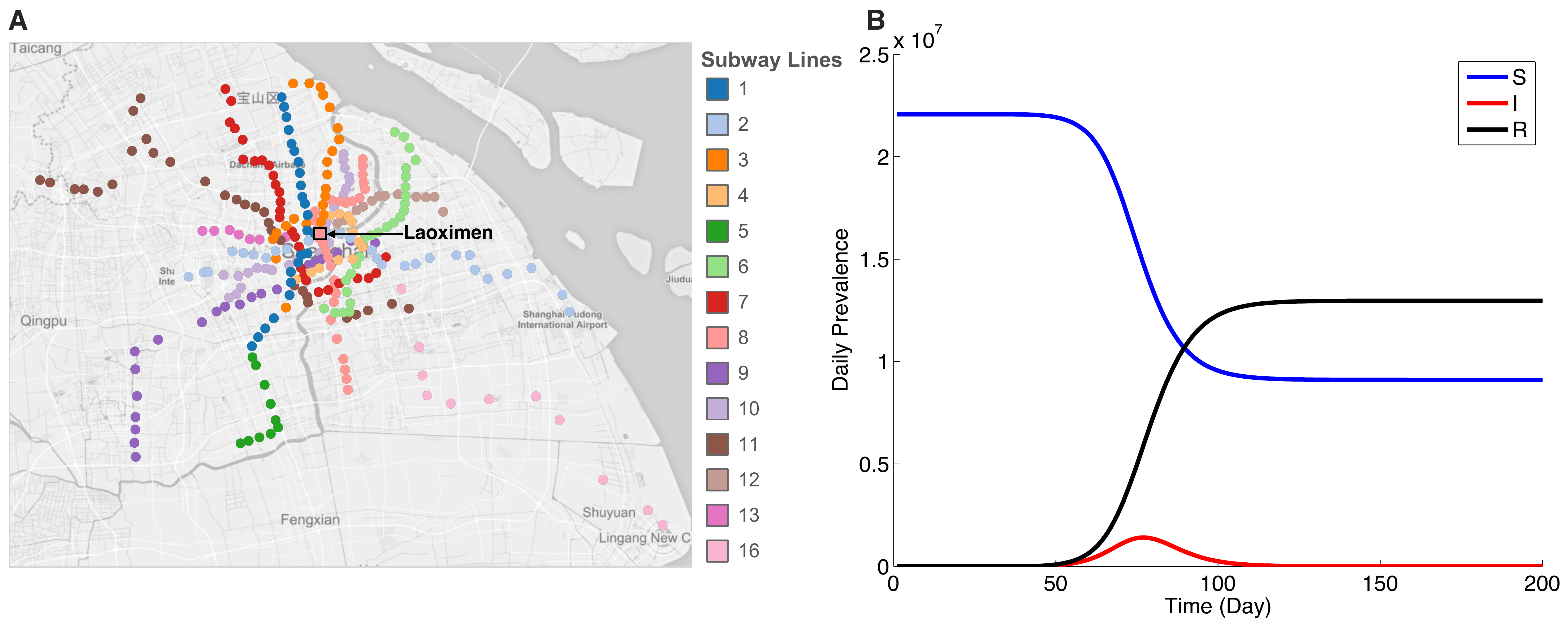} \caption{\textbf{Metapopulation susceptible-infected-recovered model for directly transmitted disease in Shanghai. }
(A) Locations (subpopulations) correspond to Shanghai subway stops. 
(B) Typical epidemic dynamics with the number of susceptible (S), infected (I) and recovered (R) individuals tracked for 200 days following the initial disease introduction. This simulation assumes that one case is introduced into the central location of Laoximen on a Monday morning with $R_0=1.5$. The epidemic runs its course in 170 days and peaks on the 77th day following the initial introduction. 
}
\label{figPlotPeakTimes} 
\end{figure}

\begin{figure}[!bpt]
\centering \includegraphics[scale=0.5]{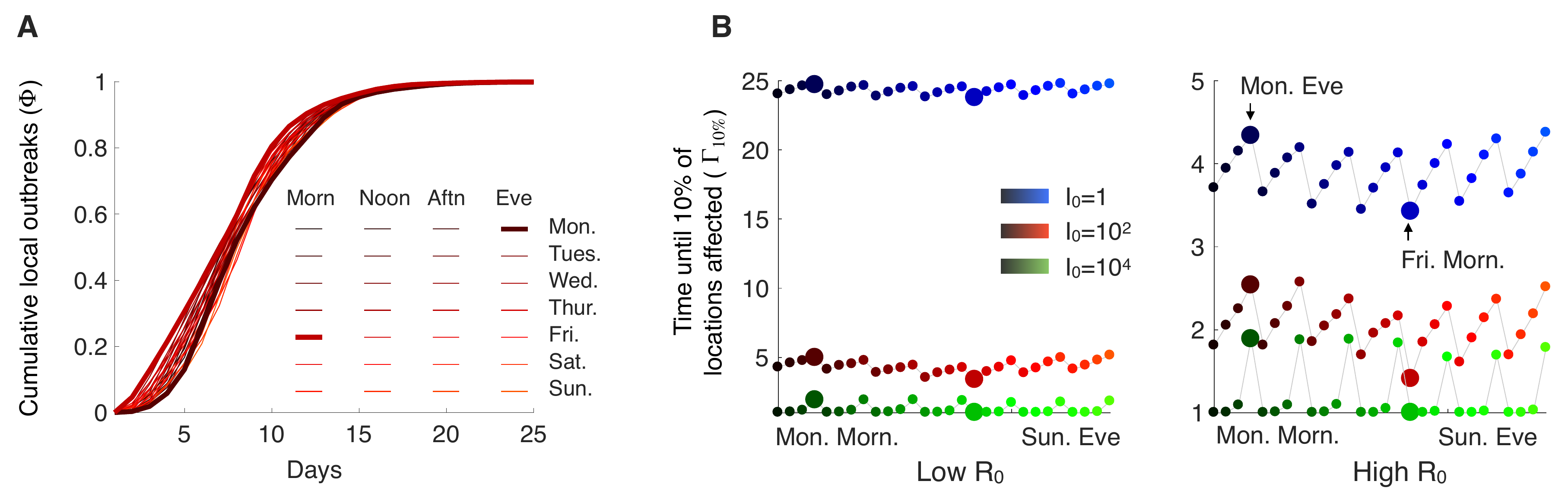} \caption{\textbf{Epidemic growth  depends on the timing of introduction.} 
(A) The cumulative fraction of new infected locations ($\Phi$), assuming $R_0=1.5$ and $I_{0}=100$. Each curve represents the average $\Phi$ across simulated epidemics starting from each of the $313$ locations. 
On average, epidemics spread fastest when they are introduced on a Friday morning and slowest from Monday evening.
(B) The time until 10\% of locations are infected ($\Gamma_{10\%}$) varies similarly across starting scenarios, with epidemics generally growing fastest when initiated in the mornings and on Friday. 
Gray lines indicate temporal sequence of starting conditions. 
}
\label{figgrowthrates} 
\end{figure}

\section*{Results}

For all combinations of introduction location and timing, we simulated epidemics using a deterministic metapopulation SIR model, assuming either low $R_0=1.5$ or high $R_0=7.5$ and initial outbreak sizes ranging from 1 to 10,000 cases, for a total of  52,584 simulations. 
For example, an epidemic introduced Monday morning at Laoximen with a single case ($I_{0}=1$) and $R_0=1.5$ eventually infects almost half of the Shanghai population, peaking on the 77-th day following the introduction and ceasing on on the 170-th day (Fig.~\ref{figPlotPeakTimes}).

\begin{figure}[!tbp]
\centering \includegraphics[scale=0.7]{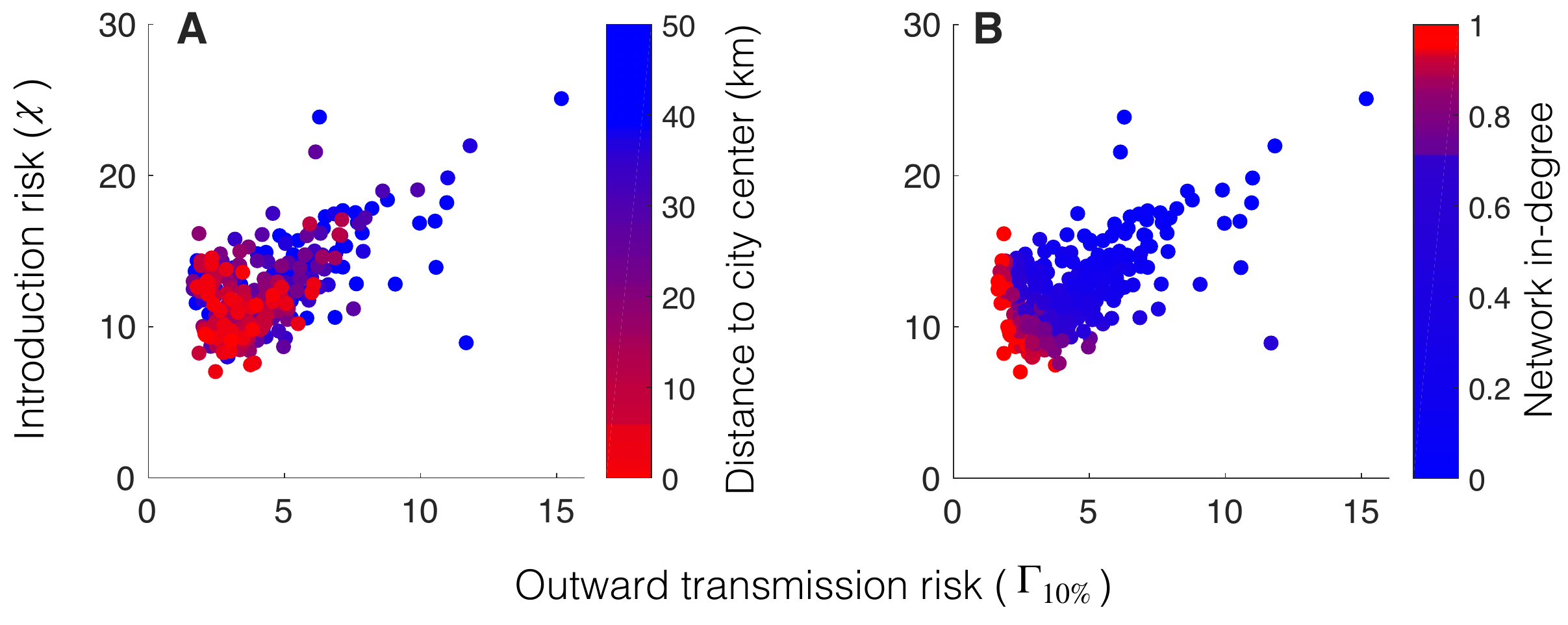} \caption{ 
\textbf{Risks of epidemic introduction and outward transmission correlate and increase towards the city center.}  
For each location, the risk of outward transmission (x-axes) is estimated by the number of days following an introduction into that location until $10\%$ of locations experience outbreaks ($\Gamma_{10\%}$), averaged over epidemics for each of the $28$ different introduction times. The risk of introduction (y-axes) is estimated by the number of days following introduction in another location until the focal location receives its first infection, averaged over all  introduction location-time combinations~($\chi$). 
Colors correspond to (A) the direct geographic distance between the location and the city center (Jing'an Temple), and (B) network in-degree (mobility flow into a location).
All simulations assume an initial outbreak of $I_{0}=100$ and $R_0=1.5$; analogous graphs for other combinations of $I_{0}$ and $R_0$ are provided in the Supplement (Fig. \ref{figCityCenterSpreadInSI}). The inward and outward transmission risks are correlated with a Pearson's correlation coefficient of 0.59, and the risks correlate to geographic centrality with coefficients of 0.44 and 0.36, and to network in-degree with coefficients of 0.63 and 0.56, respectively (all five have $p<0.05$).}
\label{figTs10} 
\end{figure}

We tracked the progression of epidemics using the cumulative fraction of new infected locations ($\Phi$). Epidemics tend to spread fastest when introduced Friday mornings and slowest from Monday evenings, with the pace of transmission intuitively increasing with both $R_0$ and $I_{0}$ (Fig.~\ref{figgrowthrates}). 
The rates at which epidemics spread \emph{from} and \emph{to} each location are highly correlated, with locations close to the city center (Jing'an Temple) tending to both spark the fastest epidemics and experience some of the earliest outbreaks (Fig.~\ref{figTs10}.A and Tab.~\ref{table3}). For $R_0=1.5$ and initial outbreak of $100$ cases, the average time between introduction and transmission to $10\%$ of all locations ranges from 2 to 15 days, and the average time until a new infection arrives in a location ranges from 7 to 25 days. The correlation coefficients between geographic distance to the city center and the inbound and outbound epidemic risks are 0.44 and 0.36, respectively (both with $p<0.05$).
We find even stronger correlations between the network centrality of a location and its epidemiological risk. For example, the flow of people into a location (in-degree) has correlation coefficients of 0.69 and 0.66 with inbound and outbound risks, respectively~(Fig.~\ref{figTs10}.B and Tab.~\ref{table3}).
These patterns hold for other combinations of $R_0$ and $I_0$ (See supplement). 
Taken together, we expect urban epidemiological risk in Shanghai to be highest for outbreaks starting on Friday mornings and in central locations, in terms of both geography and the underlying mobility network.

We elucidate the increased epidemiological risk associated with Friday morning introductions through network analysis. 
A clustering analysis of the mobility networks estimated for each of the 28 weekly time periods reveals high correlation among weekday patterns and among weekend patterns, with Friday bridging the two (Fig.~\ref{figCorr28Period}).
While the Friday morning commute resembles other weekdays, the afternoon patterns resemble the weekends.  
For every day of the week ($d$), we calculate the highest probability path between each pair of locations in Shanghai via a consecutive sequence of subway transits. From these pairwise commute probabilities, we calculate two global network statistics: the average commute probability across all pairs of locations ($\delta_d$) and network coherence ($C_d$), which is the minimum across all commute probabilities in a given day. 
By both of these measures, Friday has the highest human fluidity and thus transmission potential among all days of the week, followed by Saturday (Tab.~\ref{NetworkCoherence}). 
We map daily commute patterns to six recognizable motifs (Fig.~\ref{figActivities}). Individuals who leave home before 10:00 and return after 17:00 are classified as workers. We calculate the daily proportion of such working individuals that make additional non-work stops before returning home, and call that value the `recreational fraction' ($\rho$). The recreational fraction is higher on Fridays than other weekdays (Tab.~\ref{NetworkCoherence}).
We also calculate the afternoon influx of passengers into each location ($\sum_i F_{ij}^\text{aftn}$) and find the top ten most popular destinations for each day of the week. Many 'leisure' destinations are located close to the city center. On Monday through Thursday, there is only one location within five kilometers of the city center (Shanghai Railway Station) that ranks among the top ten; on Fridays, Saturdays and Sundays, there are four other centrally-located top ten destinations (Renmin Park, Zhongshan Park, Xujiahui, and Lujiazhui).

Finally, we partitioned the Shanghai population into communities based on transportation patterns, at both a time period a daily scale (Figs.~\ref{figCommunityAftn4P} and~\ref{figCommunityAftn}). Roughly speaking, individuals are far more likely to encounter individuals in their own community than in others. Many communities persist stably throughout out the week. However, we observe a fusing of disjoint weekday communities on Friday afternoons that persists through Saturday. For example, 71 of Shanghai's 313 locations and 26\% of the population become connected in Friday's largest community (Fig.~\ref{figCommunityAftn4P}).

\begin{table}
\centering
\caption{Weekly network dynamics in terms of coherence (minimum probability of transit between two locations), average distance (average probability of transit between two locations), and proportion of working population making non-work related stops.}
\label{NetworkCoherence}
\begin{tabular}{@{}lllll@{}}
\toprule
Day of week & Network coherence $C_d$ ($\times 10^{-9}$) & Average distance $\delta_{d}$ ($\times 10^{-4}$)    & Recreational fraction $\rho$ \\ \midrule 
Monday    & 1.66  & 2.314   & 0.2251 \\
Tuesday  & 1.66  & 2.283  & 0.2414 \\
Wednesday & 1.40  & 1.992   & 0.2222  \\
Thursday & 3.76  & 3.037   & 0.2571 \\
Friday   & 18.44  & 5.475   & 0.3323 \\
Saturday   & 16.34  & 3.293  & 0.4454  \\
Sunday   &  8.21  & 2.854    & 0.3579 \\
\bottomrule
\end{tabular}
\end{table}

\subsection*{Discussion and Conclusion}
The fate of a newly emerging infectious disease outbreak in an urban setting will likely depend on the timing and location of its initial introduction. Our analysis of a Shanghai metapopulation model suggests that the rate of epidemic expansion depends not only on well-understood  epidemiological drivers, like $R_0$ and the initial outbreak size, but also on the neighborhood of the initial cases and, to a lesser extent, the day and time they first appear. Epidemic growth rates are expected to be highest for outbreaks originating at the heart of the city, at central locations in the urban mobility network, and on Fridays. Epidemiological risk cuts both ways. Locations that tend to spark more aggressive epidemics also tend to be the earliest affected in outbreaks originating elsewhere. 

Human mobility patterns shape risk. The epidemiological importance of central neighborhoods stems primarily from connectivity rather than geography, although the two coincide.
A critical step in extrapolating this result to improve public health in other urban populations is to identify epidemiologically central neighborhoods. However, we often lack comprehensive transportation and social network data or the analytic capabilities for rigorous estimation of local risk. Fortunately, prior studies have found correlations between geographic centrality, social network centrality and population density \cite{liang2013unraveling,gallotti2015understanding}, suggesting that readily available geographic and census statistics may serve as reasonable proxies for epidemiological connectivity.

The heightened risk associated with Friday introductions likely stems from the weekday to weekend transition in urban mobility patterns. This is reflected in our network analysis of Shanghai dynamics. The likelihood that individuals will take non-work related subway trips (recreational fraction) and will travel between any given pair of locations (network coherence) peaks on Fridays and is generally higher on weekends than weekdays. Shanghai's community structure is highly dynamic on both hourly and daily scales. Despite its complexity, we observe strong partitioning that persists relatively stable from Monday through Friday morning, with extensive Friday afternoon fission and fusion bringing previously disconnected portions of the city in contact for the weekend.

Prior studies have highlighted the epidemiological importance of urban community structure~\cite{hoen2015epidemic,herrera2016disease,ball2008network}. Our findings highlight the {\it predictably} dynamic nature of urban life and consequent variation in outbreak risk, particularly that stemming from weekends, holidays, major social events and academic calendars. Based on the observed Friday effect, we hypothesize that holiday eves may generally serve as temporal hotspots for outbreak expansion.
On an international scale, connectivity between cities via ground, air and sea travel exhibits daily, weekly, seasonal and annual cycles that may similarly constrain the pace and extent of global disease emergence~\cite{belobaba2015global}.

We considered two different transmission rates and three different introduction size scenarios, ranging from a single importation to 10,000 cases appearing simultaneously. The larger initial outbreaks are motivated by plausible bioterrorist events~\cite{walden2004estimating}. For example, the 2001 anthrax release via letters in the United States initially exposed 22 people in three locations~\cite{federal2011amerithrax}; 
a 1979 accidental release of anthrax from a military microbiology facility in the city of Swerdlovsk, Russia began with 96 infections~\cite{brookmeyer2001statistical,CDC_Anthrax_2016}; in 1984, a deliberate introduction of Salmonella at ten restaurants in The Dalles, Oregon initially infected 751 people~\cite{torok1997large}. 

Urban disease introductions, whether naturally occurring or intentionally released, can rapidly expand to epidemic proportions and ignite global spread.  Surveillance and targeted early interventions may be critical to averting large scale devastation. However, not all introductions will carry the same risks. In large cities like Shanghai with complex weekly mobility patterns, outbreaks arising within highly connected neighborhoods or during time periods associated with social transitions may prove significantly more explosive that others. This study reinforces the public health importance of understanding human mobility at a high spatio-temporal resolution and the utility of incorporating predictably dynamic risk factors into epidemic prevention and response plans.


\section*{Funding}
ZW, SJF, LAM, and APG would like to acknowledge funding from the Models of Infectious Disease Agent Study (MIDAS) program grant number U01 GM087719.
PH was supported by JSPS KAKENHI grant number JP 18H01655.
The funders had no role in study design, data collection and analysis, decision to publish, or preparation of the manuscript.

\section*{Competing Interests}
The authors have declared that no competing interests exist.

\section*{Author Contributions}
Conceived and designed the experiments: ZW LAM SJF JL. Performed the experiments: ZW SJF. Analyzed the data: ZW LAM. Wrote the paper: ZW LAM PH APG.

\appendix

\begin{figure}[h]
\centering \includegraphics[scale=0.5]{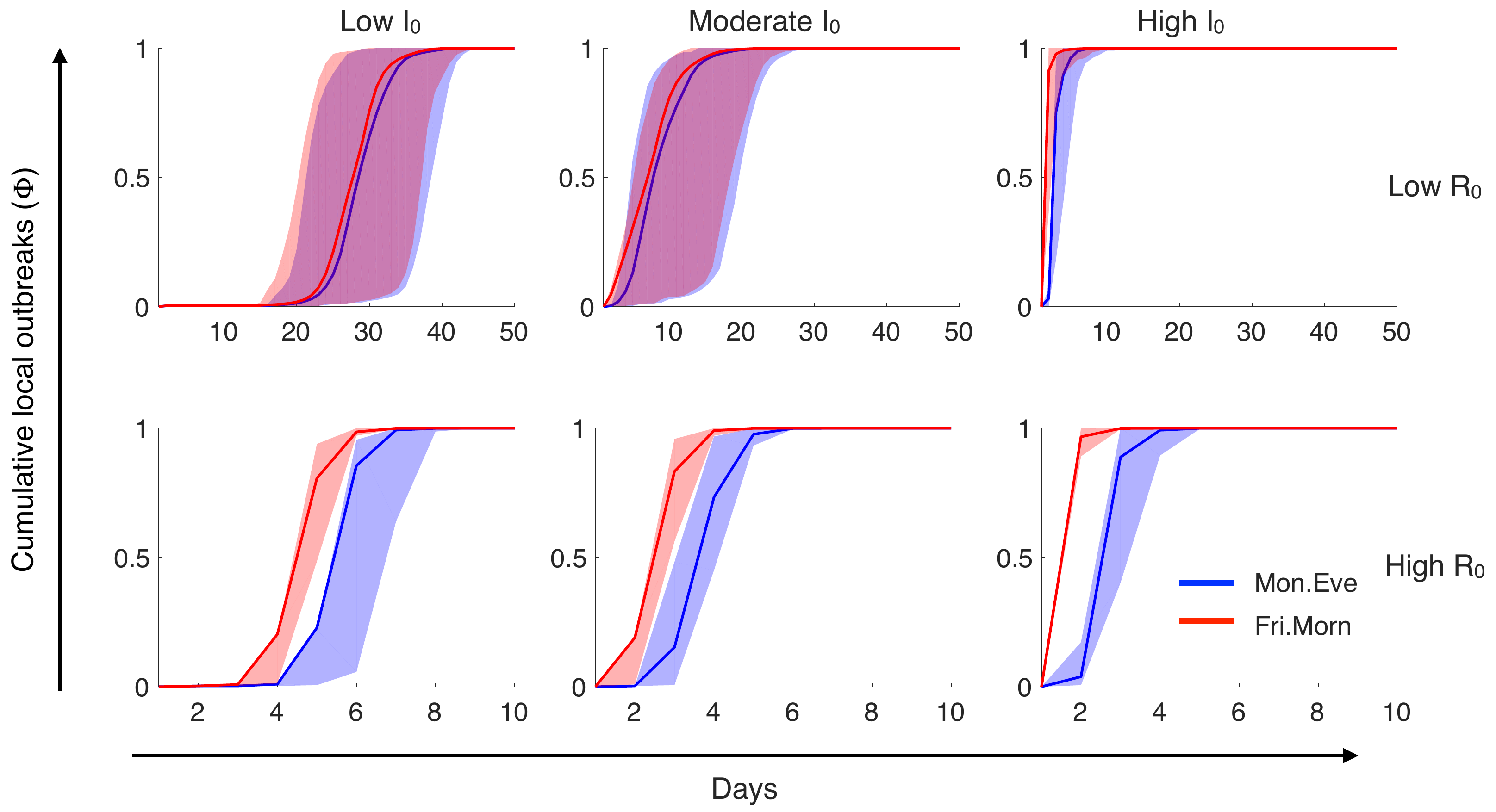} \caption{\textbf{Epidemic growth for outbreaks starting Monday evening versus Friday morning.} 
The cumulative fraction of affected locations ($\Phi$) is expected to rise most rapidly for outbreaks introduced Friday morning and most slowly for those introduced Monday evening. Lines indicate average daily cumulative $\Phi$, over all possible introduction locations; shading indicates the full range of daily cumulative $\Phi$. 
}
\label{figgrowthrates_MF} 
\end{figure}

\begin{landscape}
\begin{figure}[h]
\centering \includegraphics[scale=0.6]{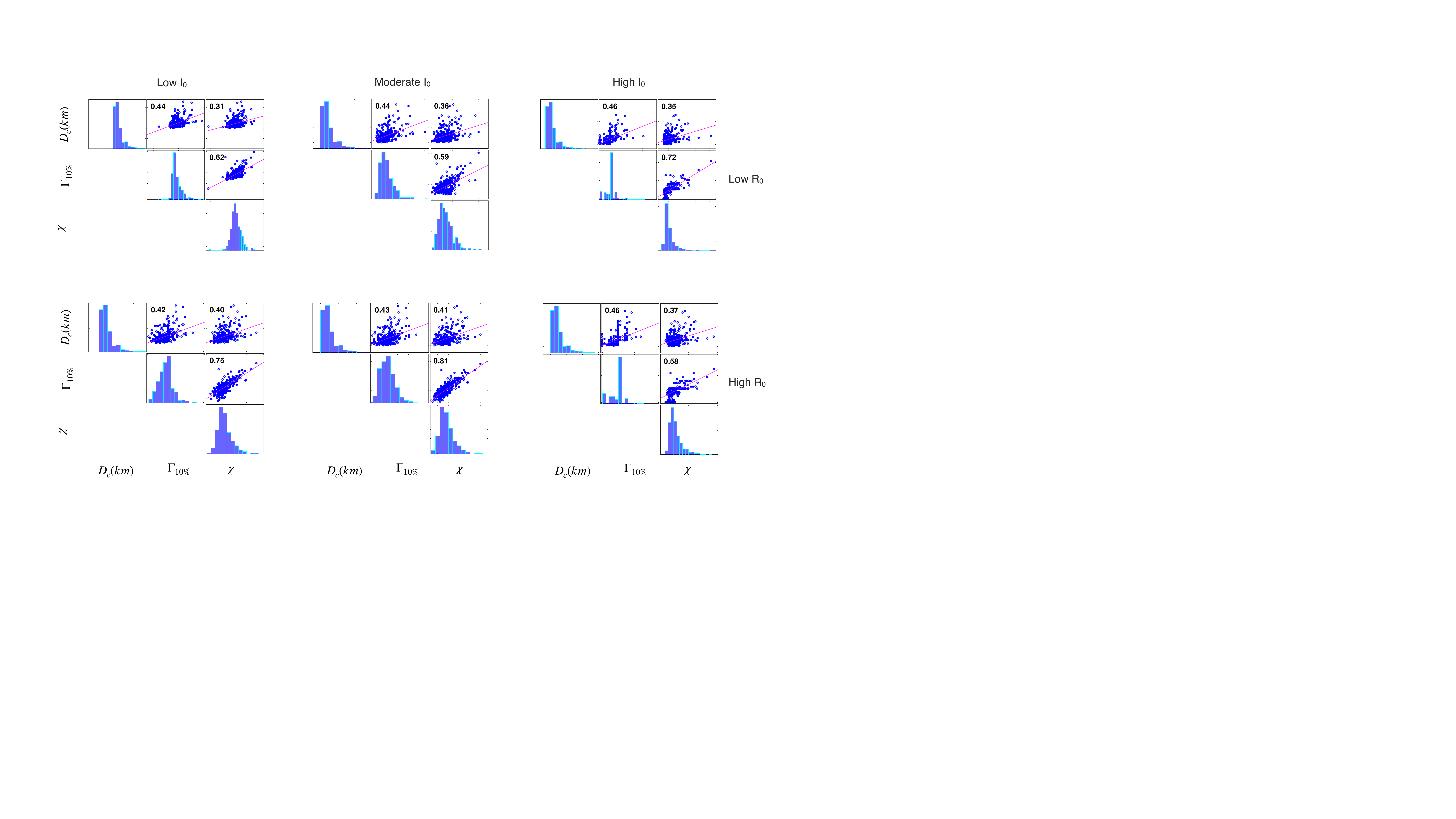} \caption{ 
\textbf{Across Shanghai's 313  locations, correlations among distance to the city center ($D_\text{c}$), speed of epidemic spread ($\Gamma_{10\%}$), and speed of epidemic arrival ($\chi$).}
Across each combination of $R_0$ (1.5 or 7) and initial outbreak size ($I_{0}$=1, 100, or 10,000), we simulated epidemics starting in each location for each of the the 28 weekly time periods. For each location (point in graph), we calculated (i) its geographic distance to the Jing'an Temple at the city center ($D_\text{c}$), (ii) the average number of days until an epidemic starting in that location reached at least $10\%$ of locations ($\Gamma_{10\%}$), across all 28 starting times, and (iii) the average number of days until epidemics starting elsewhere reached that location ($\chi$), across all combinations 312 other locations and 28 starting times.
Plots include the least-squares regression lines (red) and Pearson's correlation coefficients between each pair of metrics (all have $p < 0.05$). The strong correlations between inward and outward transmission risks increase with both $I_0$ and $R_0$. 
}
\label{figCityCenterSpreadInSI} 
\end{figure}
\end{landscape}

\begin{landscape}
\begin{figure}[h]\vspace{1.0cm}
\centering \includegraphics[scale=0.55]{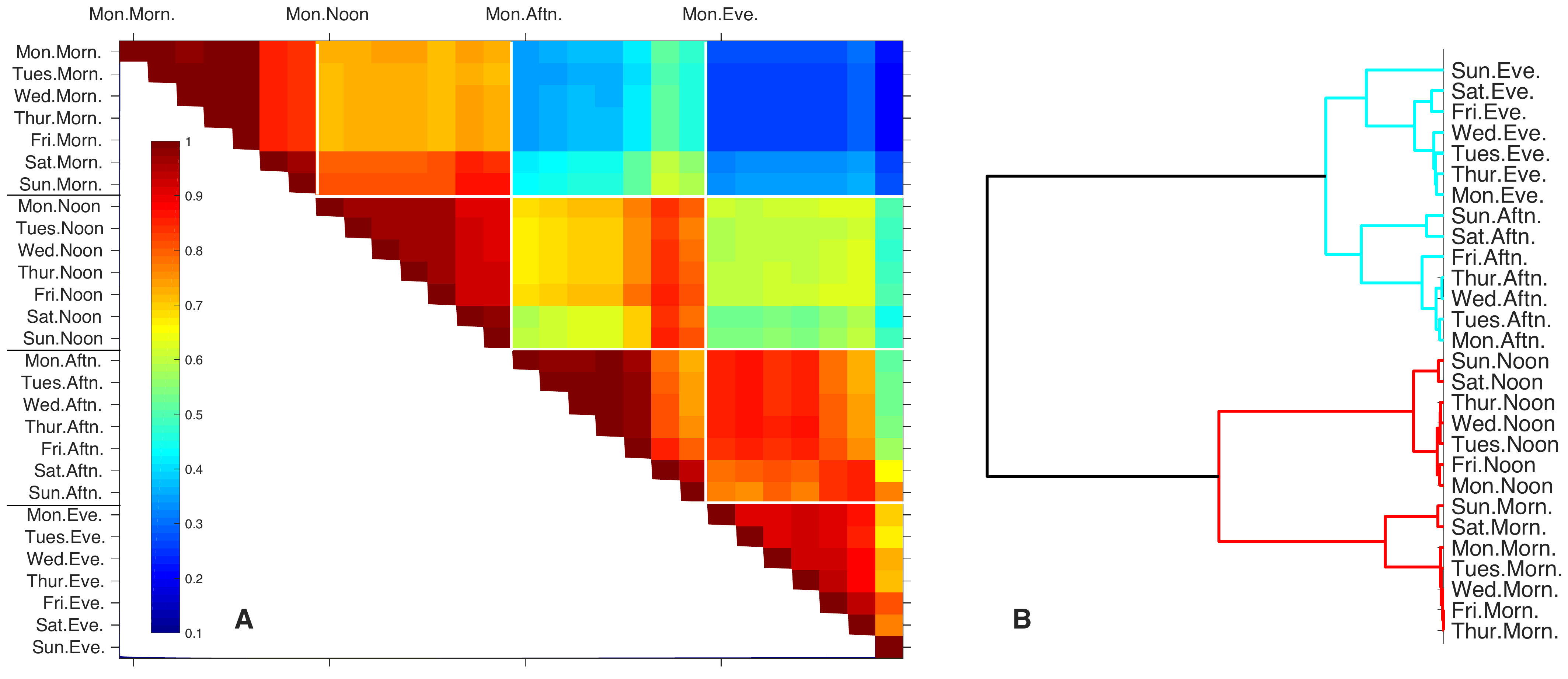} \caption{
\textbf{Correlations in mobility patterns among the 28 weekly time periods.} 
For each time period $p$, we construct the vector of all pairwise movement patterns among the 313 locations: $F^p=<F_{1,2}^p, \cdots, F_{ij}^p, \cdots, F_{313,313}^p>$  where $F_{ij}^p$  is the entry of the aggregate mobility matrix.
(A) Pearson correlations among the $F^p$ across all pairwise combinations of the 28 weekly periods. All are statistically  significant with $P < 0.05$. 
(B) Tree based on hierarchical clustering~\cite{matlab_Hierarchical} using 784 Pearson correlation coefficients among the 28 weekly periods. 
Friday morning and midday mobility patterns resemble those of other weekdays, while Friday afternoon pattern appears unique and Friday evening clusters closely with Saturday.}
\label{figCorr28Period} 
\end{figure}
\end{landscape}

\begin{figure}[h]
\centering \includegraphics[scale=0.2]{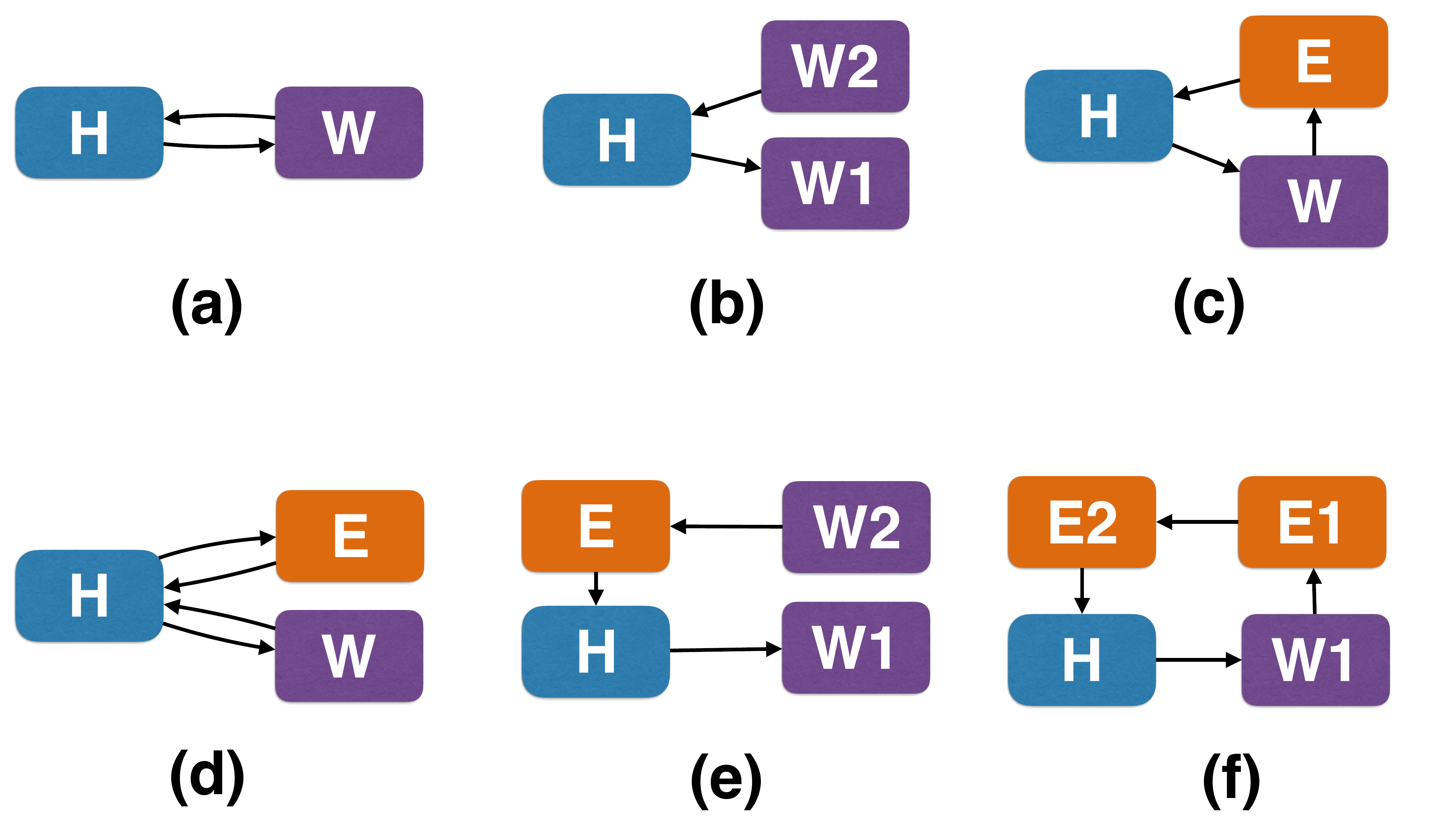} 
\caption{\textbf{Daily activity diagrams.} Workers are individuals who take their first trip of the day before 10:00 and at least one trip after 17:00. If the first origin station matches the last destination station of the day, it is designated {\it home} ($H$); the station visited for the longest period in between is designated {\it work} ($W$); all other stations visited after 16:00 are designated {\it non-work} ($E$), which may include errands, health or beauty appointments, entertainment, etc. When individuals spend time at two work or non-work locations, they are both designated accordingly ($W1$ and $W2$ or $E1$ and $E2$).
Most daily worker schedules fell into one of the six categories diagrammed above (each accounts for at least $1\%$ of the sample). 
}
\label{figActivities} 
\end{figure}

\begin{figure}[h]
\centering \includegraphics[scale=0.7]{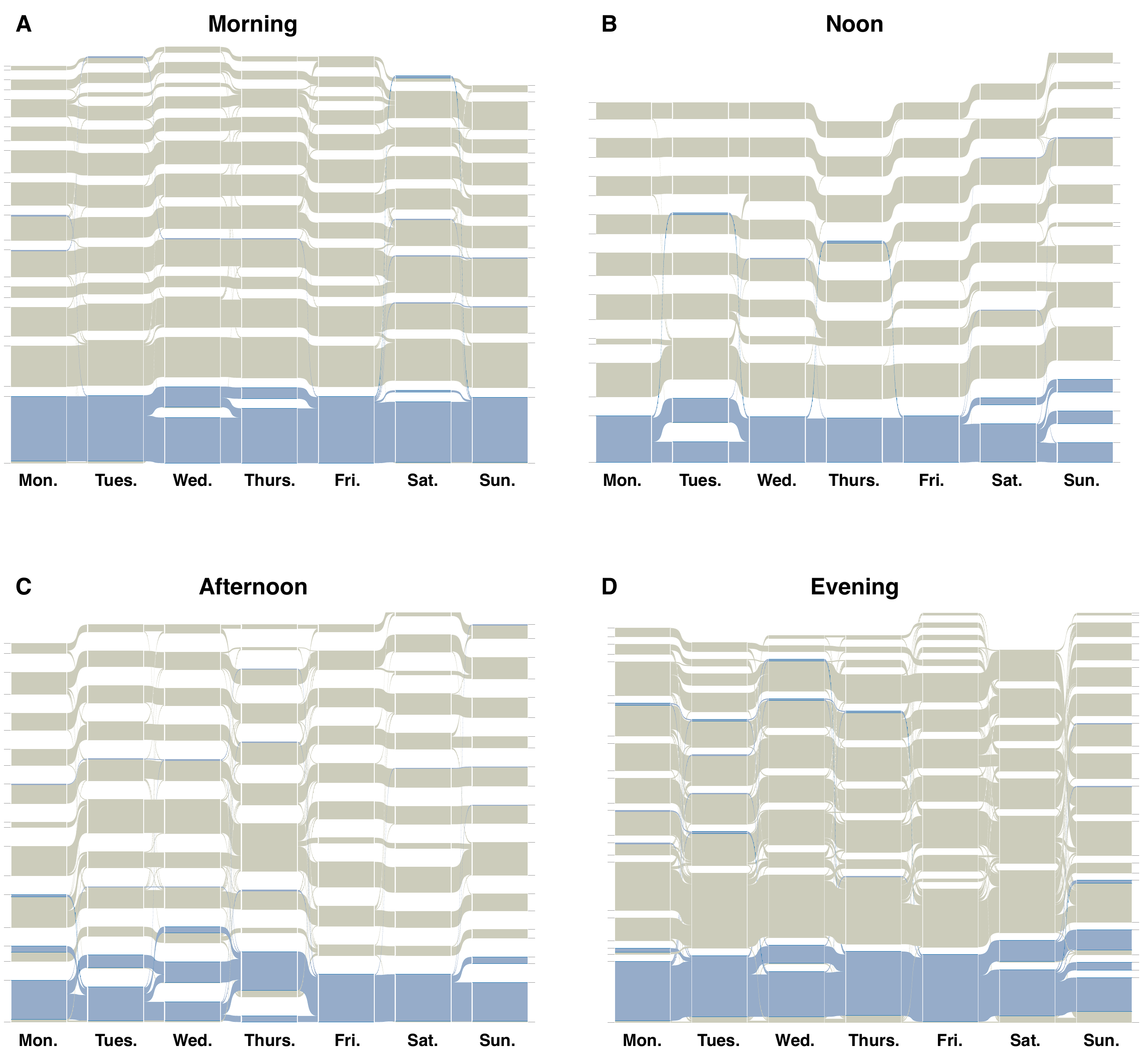} 
\caption{\textbf{Shanghai community dynamics.} 
Alluvial diagrams showing Shanghai's evolving community structure throughout the week for the (A) morning, (B) noon, (C) afternoon, and (D) evening time periods. For each of the 28 weekly time periods, we partitioned Shanghai population into highly intra-connected but loosely inter-connected communities using the Louvain community detection algorithm~\cite{blondel2008fast}. Blocks represent individual communities, with heights proportional to the number of locations included in each community. 
The height of a streamline between communities on two successive days is proportional to the number of locations that change membership from one community to the other. 
Blue indicates locations included in the largest community on Friday. While community membership is relatively stable throughout the morning, noon and evening time periods, it is considerably more dynamic during the afternoons, with previously disjoint regions of the city converging into a single community on Friday. This weekday to weekend transition from workday to recreational mobility patterns may partly explain the more rapid spread of outbreaks originating on Fridays.
}
\label{figCommunityAftn4P} 
\end{figure}

\begin{figure}[h]
\centering \includegraphics[scale=0.5]{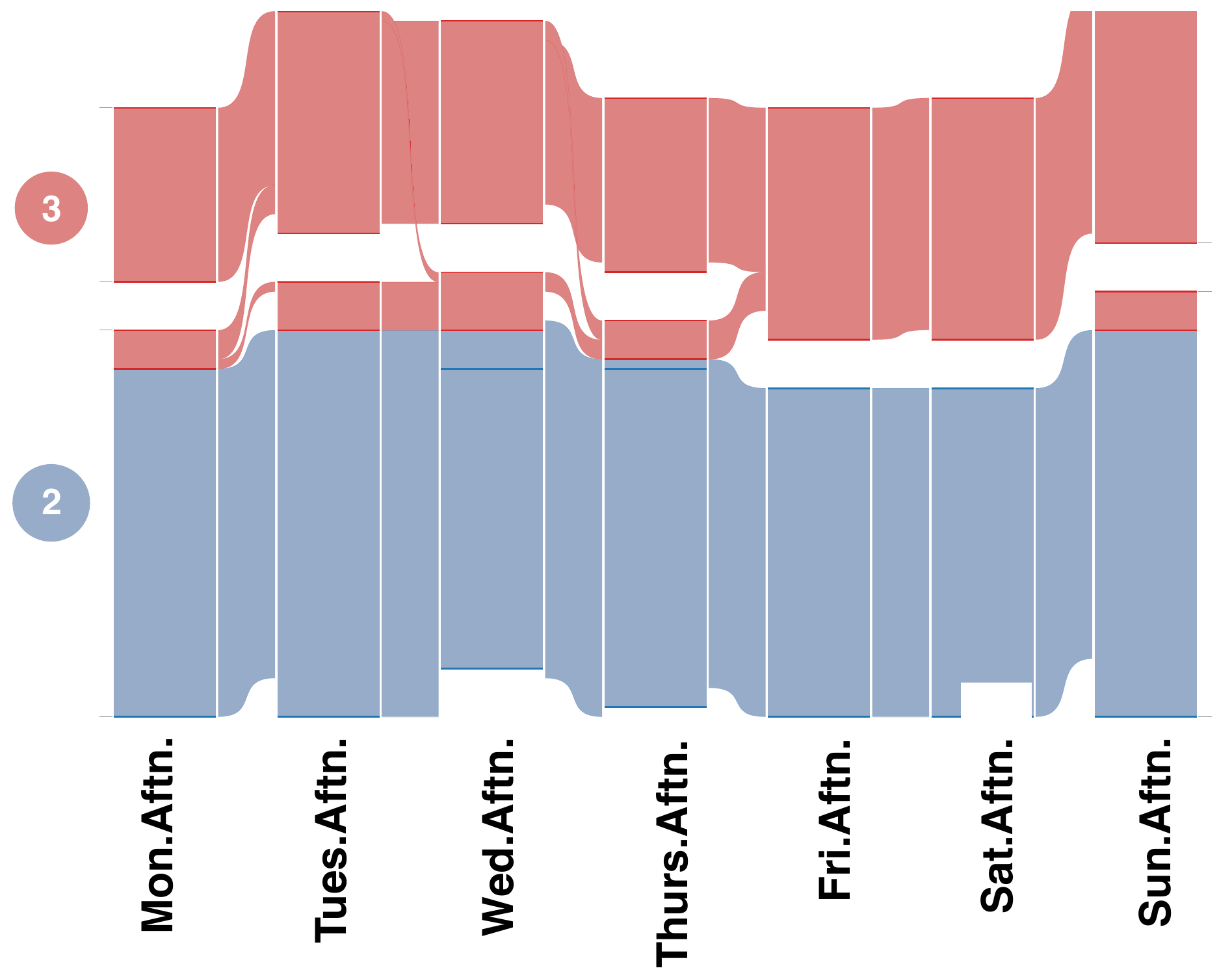} 
\caption{\textbf{Friday afternoon mixing of the second and third largest communities.} 
Alluvial diagrams for the afternoon period across the seven days of the week, considering only the second and third largest communities identified on Monday afternoon and ignoring small branches consisting of fewer than ten locations. 
For each of the 28 weekly time periods, we partitioned Shanghai population into highly intra-connected but loosely inter-connected communities using the Louvain community detection algorithm~\cite{blondel2008fast}. Blocks represent individual communities, with heights proportional to the number of locations included in each community. 
The height of a streamline between communities on two successive days is proportional to the number of locations that change membership from one community to the other. 
Red and blue reflect community membership on Friday. 
Previously disjoint red subpopulations mix on Friday and persist through Saturday.
}
\label{figCommunityAftn} 
\end{figure}

\section*{Supplement of Data Analysis}

\subsection*{Data}
We analyzed a travel smart card dataset that was compiled by Shanghai Public Transportation Card Co.Ltd for the Shanghai Open Data Apps (SODA)  contest~\cite{coltd2015}). The 30 days (April 2015) of comprehensive subway transit data includes 123 million trips taken by 11 million people among Shanghai's 313 subway stations. 
Each trip record includes (i) a smart card identification number, and the (ii) date, (iii) time, (iv) origin station, (v) destination station, and (vi) cost of the trip.
There are two periods of  peak traffic, one in the morning and one in the evening~\cite{sun2015research}. Accordingly, we divide each day into four periods (Table~\ref{table2}).
April 2015 included two holidays that likely changed mobility patterns: the three-day Qingming festival starting April 4, and the three-day Labor Day period starting May 1. 
To analyze a {\it typical} non-holiday work week, we focused our analysis exclusively on week 16.

\begin{table}[!ht]
\centering \caption{Periods involved in a day}
\label{table2} %
\begin{tabular}{ll}
\hline 
\textbf{Period Name} & \textbf{Period Section}\tabularnewline
\hline 
Morning & 5:30-10:00 a.m.\tabularnewline
Noon & 10:00-16:00\tabularnewline
Afternoon & 16:00-21:00\tabularnewline
Evening & 21:00-23:59\tabularnewline
\hline 
\end{tabular}
\end{table}

\begin{table}
\centering \caption{\textbf{Correlations among centrality measures and epidemic risks.}
The outbreak risk at a location ($\chi$) is the number days following introduction in another location until the focal location receives its first infection, averaged over all introduction location-time combinations; its outward transmission risk ($\Gamma_{10\%}$) is the number of days following an introduction into that location until $10\%$ of other locations experience outbreaks. We consider the geographic distance of the location to city center ($D_c$) and four different measures of network centrality: in-degree (direct flow into location), out-degree (direct flow out of location), in-closeness (probability of traveling directly or indirectly to the location averaged across all origins) and out-closeness (probability of traveling directly or indirectly away from the location averaged across all possible destinations). Table gives Pearson correlation coefficients among all seven epidemiological and centrality statistics for all six combinations of low and high $R_0$s and three different initial outbreak sizes ($I_0$). All correlations are statistically significant at $p<0.05$.} 
\label{table3} %
\begin{tabular}{@{}llllllllll@{}}
\toprule
$R_0$                  & $I_0$                  &                 & $\chi$ & $\Gamma_{10\%}$ & $D_c$ & in-degree & out-degree & in-closeness & out-closeness \\ \midrule
\multirow{21}{*}{1.50} & \multirow{7}{*}{1}    & $\chi$          & $1.00$   & $0.62$   & $0.31$  & $-0.46$   & $-0.11  $   & $-0.58$   & $-0.25$        \\

                      &                        & $\Gamma_{10\%}$ & $0.62$   & $1.00$   & $0.44$  & $-0.56 $   & $-0.47  $   & $-0.70$   & $-0.72 $       \\
                       &                      & $D_c$        & $0.31$   & $0.44$            & 1.00  & $-0.32 $   & $-0.20  $   & $-0.37    $   & $-0.32$        \\
                       &                        & in-degree     & $-0.46$  & $-0.56        $   & $-0.32$ & $1.00  $   & $0.77   $   & $0.86     $   & $0.78         $\\
                       &                        & out-degree    & $-0.11$  & $-0.47        $   & $-0.20$ & $0.77$     & $1.00   $   & $0.66     $   & $0.85         $\\
                       &                        & in-closeness  & $-0.58$  & $-0.70        $   & $-0.37$ & $0.86  $   & $0.66$      & $1.00     $   & $0.86         $\\
                       &                        & out-closeness  & $-0.25$  & $-0.72        $   & $-0.32$ & $0.78  $   & $0.85   $   & $0.86 $       & $1.00         $\\   \cmidrule(l){2-10} 
                       & \multirow{7}{*}{100}   & $\chi$          & $1.00$   & $0.59         $   & $0.36$  & $-0.56 $   & $-0.16  $   & $-0.77    $   & $-0.42        $\\
                       &                       & $\Gamma_{10\%}$ & $0.59$   & $1.00         $   & $0.44$  & $-0.63 $   & $-0.60  $   & $-0.77    $   & $-0.85        $\\
                       &                       & $D_c$           & $0.36$   & $0.44  $          & $1.00$  & $-0.32 $   & $-0.20  $   & $-0.37    $   & $-0.32        $\\
                       &                        & in-degree        & $-0.56$  & $-0.63        $   & $-0.32$ & $1.00  $   & $0.77   $   & $0.86     $   & $0.78         $\\
                       &                        & out-degree    & $-0.16$  & $-0.60        $   & $-0.20$ & $0.77$     & $1.00   $   & $0.66     $   & $0.85         $\\
                       &                        & in-closeness    & $-0.77$  & $-0.77        $   & $-0.37$ & $0.86  $   & $0.66 $     & $1.00     $   & $0.86         $\\
                       &                        & out-closeness  & $-0.42$  & $-0.85        $   & $-0.32$ & $0.78  $   & $0.85   $   & $0.86$        & $1.00         $\\   \cmidrule(l){2-10} 
                           & \multirow{7}{*}{10000} & $\chi$          & $1.00$   & $0.72$   & $0.35$  & $-0.56 $   & $-0.40  $   & $-0.84    $   & $-0.68        $\\
                       &                       & $\Gamma_{10\%}$ & $0.72 $  & $1.00         $   & $0.46 $ & $-0.68 $   & $-0.63  $   & $-0.77    $   & $-0.81        $\\
                       &                       & $D_c$           & $0.35$   & $0.46$            & $1.00$  & $-0.32 $   & $-0.20  $   & $-0.37    $   & $-0.32        $\\
                       &                        & in-degree        & $-0.56 $ & $-0.68        $   & $-0.32$ & $1.00  $   & $0.77   $   & $0.86     $   & $0.78         $\\
                       &                        & out-degree    & $-0.40$  & $-0.63        $   & $-0.20$ & $0.77$     & $1.00   $   & $0.66     $   & $0.85         $\\
                       &                        & in-closeness    & $-0.84$  & $-0.77        $   & $-0.37$ & $0.86  $   & $0.66$      & $1.00     $   & $0.86         $\\
                       &                        & out-closeness    & $-0.68$  & $-0.81        $   & $-0.32$ & $0.78  $   & $0.85   $   & $0.86$        & $1.00         $\\       \bottomrule
\multirow{21}{*}{7.50}&\multirow{7}{*}{1}&$\chi$&$1.00$&$0.75$&$0.40$&$-0.72$&$-0.43$&$-0.92$&$-0.69$\\
&&$\Gamma_{10\%}$&$0.75$&$1.00$&$0.42$&$-0.73$&$-0.74$&$-0.80$&$-0.92$\\
&&$D_c$&$0.40$&$0.42$&$1.00$&$-0.32$&$-0.20$&$-0.37$&$-0.32$\\
&&in-degree&$-0.72$&$-0.73$&$-0.32$&$1.00$&$0.77$&$0.86$&$0.78$\\
&&out-degree&$-0.43$&$-0.74$&$-0.20$&$0.77$&$1.00$&$0.66$&$0.85$\\
&&in-closeness&$-0.92$&$-0.80$&$-0.37$&$0.86$&$0.66$&$1.00$&$0.86$\\
&&out-closeness&$-0.69$&$-0.92$&$-0.32$&$0.78$&$0.85$&$0.86$&$1.00$\\\cmidrule(l){2-10}
&\multirow{7}{*}{100}&$\chi$&$1.00$&$0.81$&$0.41$&$-0.75$&$-0.52$&$-0.95$&$-0.77$\\
&&$\Gamma_{10\%}$&$0.81$&$1.00$&$0.43$&$-0.72$&$-0.71$&$-0.81$&$-0.92$\\
&&$D_c$&$0.41$&$0.43$&$1.00$&$-0.32$&$-0.20$&$-0.37$&$-0.32$\\
&&in-degree&$-0.75$&$-0.72$&$-0.32$&$1.00$&$0.77$&$0.86$&$0.78$\\
&&out-degree&$-0.52$&$-0.71$&$-0.20$&$0.77$&$1.00$&$0.66$&$0.85$\\
&&in-closeness&$-0.95$&$-0.81$&$-0.37$&$0.86$&$0.66$&$1.00$&$0.86$\\
&&out-closeness&$-0.77$&$-0.92$&$-0.32$&$0.78$&$0.85$&$0.86$&$1.00$\\\cmidrule(l){2-10}
&\multirow{7}{*}{10000}&$\chi$&$1.00$&$0.58$&$0.37$&$-0.62$&$-0.46$&$-0.88$&$-0.73$\\
&&$\Gamma_{10\%}$&$0.58$&$1.00$&$0.46$&$-0.72$&$-0.71$&$-0.73$&$-0.83$\\
&&$D_c$&$0.37$&$0.46$&$1.00$&$-0.32$&$-0.20$&$-0.37$&$-0.32$\\
&&in-degree&$-0.62$&$-0.72$&$-0.32$&$1.00$&$0.77$&$0.86$&$0.78$\\
&&out-degree&$-0.46$&$-0.71$&$-0.20$&$0.77$&$1.00$&$0.66$&$0.85$\\
&&in-closeness&$-0.88$&$-0.73$&$-0.37$&$0.86$&$0.66$&$1.00$&$0.86$\\
&&out-closeness&$-0.73$&$-0.83$&$-0.32$&$0.78$&$0.85$&$0.86$&$1.00$\\\bottomrule
\end{tabular}
\end{table}

\end{document}